\documentclass[12pt]{iopart}

\usepackage{iopams}
\usepackage{graphicx}
\newcommand{\onlinecite}{\cite}
\newcommand{\Rvec}{\boldsymbol{R}}
\newcommand{\rvec}{\boldsymbol{r}}
\newcommand{\kvec}{\boldsymbol{k}}
\newcommand{\uvec}{\boldsymbol{u}}

\newcommand{\nvec}{\boldsymbol{n}}

\newcommand{\xivec}{\boldsymbol{\xi}}

\begin{document}

\title[Quantum simulation of strongly deformable materials]{Quantum simulation 
of electron-phonon interactions in strongly deformable materials.}


\author{J.P. Hague and C. MacCormick}
\address{Department of Physical Sciences, The Open University, Walton Hall, Milton Keynes, MK7 6AA, UK}

\begin{abstract}
We propose an approach for quantum simulation of electron-phonon
interactions using Rydberg states of cold atoms and ions. We show
how systems of cold atoms and ions can be mapped onto electron-phonon
systems of the Su--Schrieffer--Heeger type. We discuss how properties
of the simulated Hamiltonian can be tuned and how to read physically
relevant properties from the simulator. In particular, use of painted
spot potentials offers a high level of tunability, enabling all
physically relevant regimes of the electron-phonon Hamiltonian to be
accessed.
\end{abstract}

\date{5th September 2011}
\pacs{37.10.Jk, 37.10.Ty, 73.20.Mf, 32.80.Ee, 32.80.Qk}
 


\section{Introduction}

Electron-phonon interactions lead to some dramatic effects in
condensed matter systems, with lattice vibrations leading to phenomena
such as superconductivity \cite{bardeen1957a} and colossal
magnetoresistance \cite{roennow2005a}. Even in the presence of
overwhelming Coulomb repulsion, quasi-particle properties such as the
effective mass can be strongly modified when electron-phonon
interactions are strong. The importance of electron-phonon
interactions has led to an industry in numerical simulation, but even
very simplified models \cite{holstein1959a,su1980a} are significantly
more difficult to treat than other standard models in condensed matter
such as Hubbard's model \cite{hubbard1963a}. This difficulty lies in
the very large Hilbert space that is needed to handle a potentially
infinite number of phonons associated even with motion of a single electron.

Recently, there has been a move to use systems of cold atoms as
quantum simulators to investigate standard models of condensed matter
physics \cite{bloch2008a}. The model that has been most successfully
simulated is the Hubbard model of instantaneous local repulsion
between electrons \cite{hubbard1963a}. An optical lattice loaded from a Bose-Einstein condensate can be directly related to a Hubbard model, since there is a
large energy penalty for two atoms to sit on the same lattice site
\cite{bloch2008a}. The use of cold atoms to simulate Hubbard models
has led to direct observation of important phenomena such as the
superfluid to Mott insulator transition
\cite{greiner2002a,campbell2006a}. However, there is no simple way of
extending the approach to include phonons.

There are several classes of electron-phonon interaction that are of
direct relevance to condensed matter systems.  For polymers, which are
very easy to deform, the hopping of electrons along the chain can be
strongly modified by the presence of lattice vibrations. This leads to
the Su--Schrieffer--Heeger (SSH) model, which describes a chain of CH
groups in polyacetylene \cite{su1980a}, which for a chain of $N$ groups
of mass $M$ has the Hamiltonian,
\begin{equation}
H_{{\rm SSH}}  =  -\sum_{i\sigma}t_{i+1,i}(c^{\dagger}_{i+1,\sigma}c_{i,\sigma}+{\rm H.c.})
+ \frac{K}{2}\sum_{i}(u_{i+1}-u_{i})^2  + \frac{M}{2}\sum_i \dot{u}_i^2,
\label{eqn:ssh}
\end{equation}
where the intersite hopping, $t_{i+1,i} = t_0-g(u_{i+1}-u_{i})$,
$c^{\dagger}_i$ creates an electron at site $i$, $g$ is the
electron-phonon coupling strength, $t_0$ is the hopping in the absence
of vibrations, $\sigma$ is the electron spin, $u_{i}$ is the group
displacement and $K$ is the spring constant
for the phonons. The first term in this Hamiltonian is the electron
kinetic energy from hopping, the second term is the potential energy
from bond stretching and the third term the kinetic energy of the
nuclei. In such strongly deformable materials, the vibrational
amplitude of the atoms is large enough to affect the hopping
integrals, since electrons typically hop more easily between atoms
that are closer together. Thus the hopping, $t_0$, is augmented
proportional to the relative displacement of neighbouring atoms. The
energy spectrum for the second and third terms in the Hamiltonian, that
relate to the phonon modes, may be solved in the absence of
interaction by Fourier transforming, diagonalising and introducing
creation and annihilation operators \cite{march1967a}. In this way,
they are rewritten as
$\sum_{\kvec\nu}\hbar\omega_{\kvec\nu}(d^{\dagger}_{\kvec\nu}d_{\kvec\nu}+1/2)$,
where each phonon mode has energy $\hbar \omega_{\kvec\nu}$. The
combination $d^{\dagger}_{\kvec\nu} d_{\kvec\nu}$ is the number
operator for phonons of mode $\nu$ with momentum $\kvec$, where
$d^{\dagger}_{\kvec\nu}$ and $d_{\kvec\nu}$ respectively represent
phonon creation and annihilation operators.

General three-dimensional phonon fields are quantized in the usual way
by substituting
\begin{equation}
\uvec_{i}=\sum_{\kvec,\nu}\sqrt{\frac{\hbar}{2NM\omega_{\kvec\nu}}}\xivec_{\kvec\nu}\left(d_{\kvec\nu}e^{-i\kvec\cdot\Rvec_i}+d^{\dagger}_{\kvec\nu}e^{i\kvec\cdot\Rvec_i}\right).
\label{eqn:quantise}
\end{equation}
with $\xivec_{\kvec\nu}$ the polarisation vector \cite{march1967a}.
In the case of the standard SSH model, there is a single phonon mode,
so the model is simplified and
$u_i=\sum_{\kvec}\sqrt{\hbar/2NM\omega_{\kvec}}\left(d_{\kvec}e^{-i\kvec\cdot\Rvec_i}+d^{\dagger}_{\kvec}e^{i\kvec\cdot\Rvec_i}\right)$. Creation
and annihilation operators may be Fourier transformed using the
relation,
$d^{\dagger}_{i}=\sum_{\kvec}d^{\dagger}_{\kvec}e^{i\kvec.\Rvec_i}/\sqrt{N}$,
where $d^{\dagger}_i$ creates a phonon at site $i$.  If the phonons
are taken to be a dispersionless Einstein modes of frequency
$\omega_0$ (which are a mean field approximation to the phonons, and
are often a good approximation to optical phonon modes) then the
momentum sum can be performed to obtain the
expression,
$u_i=\sqrt{\hbar/2M\omega_{0}}\left(d_{i}+d^{\dagger}_{i}\right)$. In
the case of polymers, out of chain phonon modes may also be
permissible, and if Einstein phonons are used as an approximation, the
polarization vectors $\xivec_{\kvec\nu}$ point in orthogonal
directions.

There are a number of other models of electron-phonon interactions in condensed matter. One of the most prominent is Holstein's molecular
crystal model, which describes the type of interactions that can be found in
many materials, where the local electron density couples to local
optical phonon modes which are usually associated with vibrations of
other ions in the unit cell \cite{holstein1959a}. The Fr\"ohlich model
describes the continuum limit of interactions between electrons and
strongly polarisable materials, which originate when itinerant electrons
generate a dipole moment in a medium \cite{frohlich1952a} and can
be generalised to lattice models \cite{alexandrov1999a} (the latter
are sometimes known as extended Holstein interactions). Holstein and
extended Holstein interactions have the form,
\begin{equation}
H_{\rm Holstein} = -t_0\sum_{\langle ij\rangle\sigma}c^{\dagger}_{i\sigma}c_{j\sigma}+\sum_{ij\sigma} \bar{g}_{ij} n_{i\sigma}(d^{\dagger}_{j}+d_{j})+\hbar\omega_0 \sum_{j} \left( d^{\dagger}_{j}d_{j} + 1/2 \right)
\label{eqn:holstein}
\end{equation}
where $\bar{g}_{ij}$ is related to the force between the electron and
the ion, and for a standard Holstein model
$\bar{g}_{ij}=\bar{g}\delta_{ij}$, where $\delta_{ij}$ is the
Kronecker delta. The local number operator for electrons of spin
$\sigma$ at site $i$ is $n_{i\sigma} =
c^{\dagger}_{i\sigma}c_{i\sigma}$, and $d^{\dagger}_j d_j$ is the
number operator for phonons at site $j$. The third term in the
Hamiltonian represents a local Harmonic oscillator or Einstein
mode. Since $d^{\dagger}_j+d_j\propto u_j$ where $u_j$ is the atomic
displacement, a simple physical picture arises: when an electron hops
onto a site, an atomic displacement is formed, and as the electron
leaves a site, the displacement relaxes. The continual displacement
and relaxation of atoms generates a phonon cloud that
moves with the electrons, increasing their effective mass.

Even the simplest case of the electron-phonon model (the polaron)
where a single electron moves on a chain has caused controversy since
the polaron concept was introduced \cite{landau1933a}. In particular,
the existence of self-trapping, where distortions of the lattice
generated by the electron prevent the electron from hopping, has
remained an open question. It is now known that self-trapping is not
possible in the Holstein model, where the effective mass increases
exponentially at strong coupling, but never becomes infinite
\cite{kornilovitch1998a}. On the other hand, there is tantilising
evidence for self-trapping in the SSH model which may have
consequences for polymer physics \cite{marchand2010a}.

A number of schemes for quantum simulation of different interaction
types between electrons and phonons have been suggested. A scheme for
simulation of the Frenkel--Kontorova model to probe friction and
energy transport has been proposed by Pruttivarasin {\it et al.}
\cite{pruttivarasin2011a}. Bruderer {\it et al.} have proposed a
system where an optical lattice is bathed in a BEC
\cite{bruderer2007a, bruderer2010a} in which polaron effects have been
observed \cite{gadway2010a}. The use of cold polar molecules to obtain
Holstein polaron effects has been discussed by Herrera {\it et al.} \cite{herrera2011a}. It is also appropriate to note the use of
Rydberg ions to simulate spin systems, where high-energy phonon states
are used as part of the mapping onto a spin system. Porras and Cirac
have made the mapping between a system of ions and an Ising model by
making a Lang--Firsov transformation \cite{porras2004a}. Their
proposed system has the form of phonons coupled to a single Rydberg
state via an interaction of the Holstein type, and the interaction is
generated using a laser. M\"uller {\it et al.} have proposed the use
of trapped Rydberg ions to simulate spin chains
\cite{mueller2008a}. Li and Lesanovsky have discussed the structural
distortions associated with exciting a high-energy Rydberg state in a
cold ion crystal \cite{li2011a}. We also note the proposed use of
weakly dressed Rydberg states to simulate supersolids
\cite{henkel2010a}, strongly correlated gases \cite{pupillo2010a} and
collective many-body interactions \cite{honer2010a}.

In this article, we present an approach to simulating the SSH polaron
problem using Rydberg states of cold atoms and ions. We will show how
a system of Rydberg atoms can be directly mapped to a
Su--Schrieffer--Heeger electron-phonon Hamiltonian, which is of direct
interest in the modelling of polymers and other condensed matter
phenomena relating to highly deformable materials, and to our
knowledge has not been investigated before in the context of cold
atoms. The simulator is introduced and the mapping to corresponding
models is carried out. We then discuss how to read properties of
electron-phonon interactions from the simulator, and describe a simple
experiment to investigate the self-trapping problem.

\begin{figure}
\begin{indented}
\item[]\includegraphics[width=100mm]{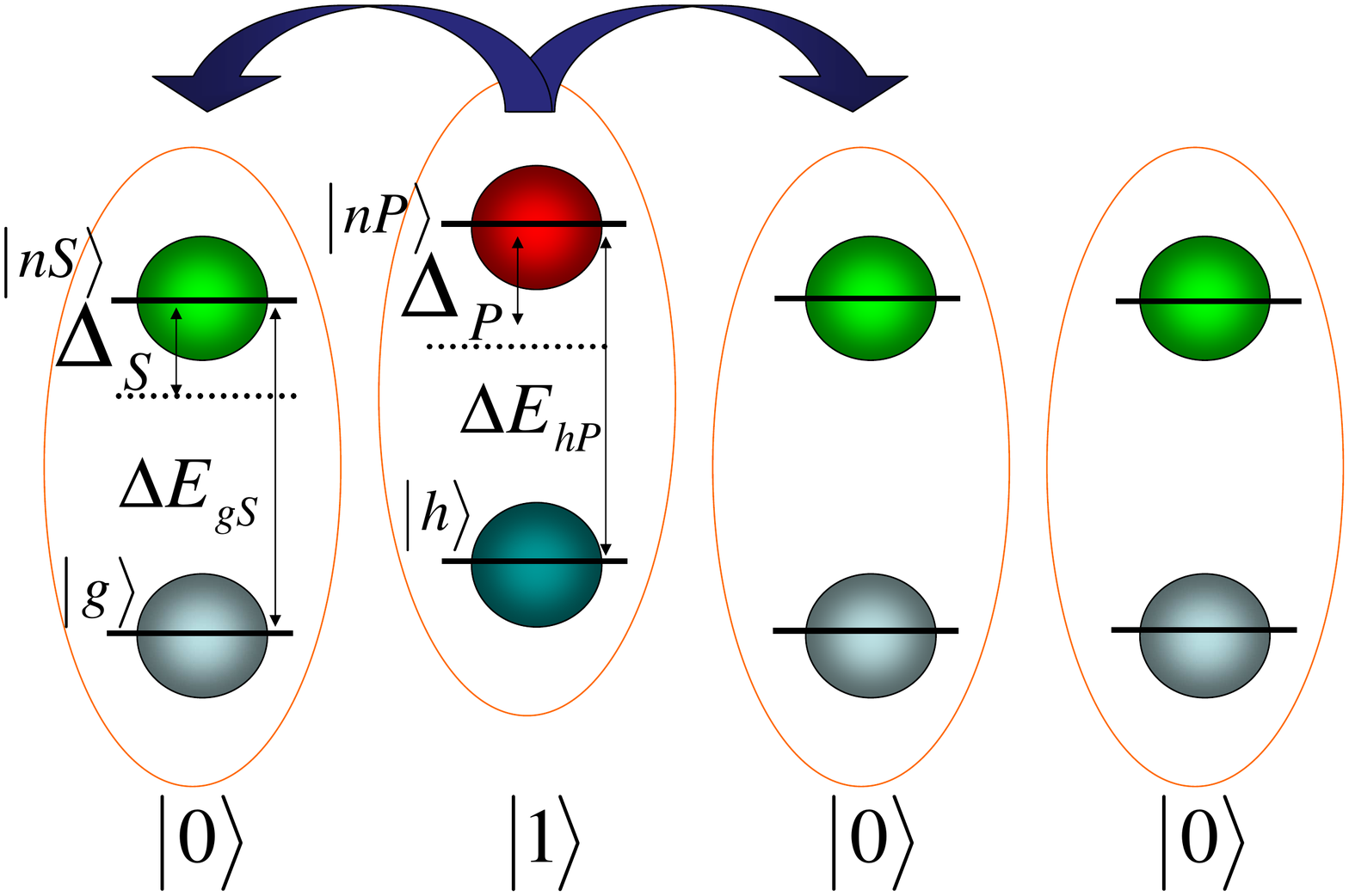}
\end{indented}
\caption{Schematic of the dressed Rydberg excitations. A set of cold
  atoms or ions is prepared. They are illuminated by two frequencies $\Delta E_{gS}-\Delta_{S}$ and $\Delta E_{hP}-\Delta_{P}$, where $\Delta E_{gS}$ corresponds to the transition between the $|g\rangle$ and $|nS\rangle$ states, and $\Delta E_{hP}$ to the transition between $|h\rangle$ and $|nP\rangle$ states. $|g\rangle$ and $|h\rangle$ are hyperfine levels and $|nS\rangle$ and $|nP\rangle$ Rydberg states with $l=0$ and $l=1$ respectively, both with $m_l=0$. The lasers are de-tuned from the transition so that the particles spend only a short period of time in the Rydberg states. This increases the lifetime of the particles by reducing the incidence of ionization, and also stabilises the particles against decay, although this reduces the effective dipole-dipole interactions between the atoms. Superpositions of the $|h\rangle$ and $|nP\rangle$ states move around the lattice as excitons.}
\label{fig:rydbergschema}
\end{figure}

\section{Mapping between cold Rydberg atoms and a Su--Schrieffer--Heeger model}
\label{sec:coldatoms}

We now consider the various terms in the Hamiltonian that describes
the motion of excited Rydberg states, the interaction of the Rydberg
state with ions and atoms, and the interaction of atoms and ions with
each other.  We begin by discussing how electron-phonon interactions
can be simulated in a system of cold Rydberg atoms such as rubidium in
a deep optical lattice where there is a single atom per lattice
site. The deep optical lattice is essential in this case so that
hoppings of atoms between sites are suppressed and there is no double
occupancy of atoms. Systems of cold ions will be discussed later.

Following the prescription in Ref. \onlinecite{wuester2011a}, it is
possible to set up a stable (long lifetime) Rydberg system where the
excitations are superpositions of low-energy hyperfine and high-energy
Rydberg states. Such a system is initialized with 2 hyperfine states
$|g\rangle$ and $|h\rangle$, so that (for example) on a chain the system is in
a state $|\pi_i\rangle = |\cdots ggghggg\cdots\rangle$. Then two
lasers detuned from the $|g\rangle\leftrightarrow|nS\rangle$ and
$|h\rangle\leftrightarrow|nP\rangle$ transitions of the cold atoms
from the ground to Rydberg states $|nP\rangle$ and $|nS\rangle$ are
switched on so that all but one atom is in the dressed
state $|0\rangle=c_S|nS\rangle+c_g|g\rangle$. Excitons in state
$|1\rangle=c_P|nP\rangle+c_h|h\rangle$ hop in the system, as
illustrated in figure \ref{fig:rydbergschema}. Here $c_S, c_g, c_P$
and $c_h$ are constants. As described in Ref. \cite{wuester2011a} the
van Vleck perturbation theory (which acts to project an interacting
Hamiltonian onto an effective hopping Hamiltonian) can be used to
obtain an effective Hamiltonian for the motion of excitons,
\begin{equation}
H_{\rm eff} = \sum_{i}\left(E_2 + E_4+ W\right)|\tilde{\pi}_i\rangle\langle \tilde{\pi}_i| + \sum_{i\neq j}\tilde{V}_{ij} |\tilde{\pi}_i\rangle\langle \tilde{\pi}_j| \label{eqn:effhamiltonian}
\end{equation}
where the wavefunction $|\tilde{\pi}_i\rangle$ represents a single exciton at site $i$. There are onsite energy offsets as a direct result of the dressing of the Rydberg states, which are calculated to 2nd order as,
\begin{equation}
E_2 = (N-1)\alpha_{S}^{2}\Delta_S + \alpha_P^2\Delta_P
\end{equation}
and to 4th order in the van Vleck scheme as,
\begin{equation}
E_4 = -(N-1)\alpha_S^4\Delta_S + \alpha_P^4\Delta_P-(N-1)\alpha^2_S\alpha_P^2(\Delta_S+\Delta_P)]
\end{equation}
and there is an additional energy shift due to dipole-dipole interactions of,
\begin{equation}
W = \alpha_S^2 \alpha_P^2\left(\sum_{k\neq j}\frac{1}{1-\bar{V}_{kj}^2}\right)(\Delta_S+\Delta_P)
\label{eqn:dipoledressedshift}
\end{equation}
A hopping term between exciton states on sites $i$ and $j$ also emerges from the perturbation theory,
\begin{equation}
\tilde{V}_{ij} = \alpha_S^2\alpha_P^2\frac{V_{ij}}{1-\bar{V}_{ij}^2}
\label{eqn:dipoledressed}
\end{equation}
where $\bar{V}_{ij} = V_{ij} / (\Delta_S + \Delta_P)$. $N$ is the
number of ions in the chain. $\alpha_S=\Omega_S/2\Delta_S$ and
$\alpha_P=\Omega_P/2\Delta_P$ are dressing parameters. $\Delta_S$ and
$\Delta_P$ are the detunings of the illuminating radiation away from
the frequencies corresponding to the $g$-$nS$ gap, $\Delta E_{gS}$, and
the $h$-$nP$ gap, $\Delta E_{hP}$, respectively. The Rabi frequencies are
denoted $\Omega_S$ and $\Omega_P$. The undressed interaction between
Rydberg states on different ions is given by the dipole-dipole interaction,
\begin{equation}
V_{kl} = -\frac{\mu^2}{|\Rvec_k -\Rvec_l|^3},
\label{eqn:dipdipint}
\end{equation}
where $\mu$ is the effective dipole moment on the Rydberg atom and $\Rvec_{k}$ is the vector to the $k$th atom in the optical lattice.

All states are set up with $m_l=0$, and the atoms are illuminated with
light which is polarized out of the plane, fixing the dipole moments
perpendicular to the interatomic vectors. With this choice of laser
polarisation, transitions to the $|nP\rangle, m_l=1,-1$ states are
suppressed \cite{moebius2011a}. A further utility of this set-up will
become apparent when cold ions are discussed. It is worth noting that
the effective Hamiltonian can also be written using the second
quantised notation as,
  \begin{equation}
H_{\rm eff} = \sum_{i}\left(E_2 + E_4+W\right)a^{\dagger}_i a_i + \sum_{i\neq j}\tilde{V}_{ij} a^{\dagger}_i a_j \label{eqn:effhamiltonianquantized}
\end{equation}
where $a^{\dagger}_i$ creates an exciton and $a_i$ annihilates an
exciton at site $i$. In the following, the excitons will take the
place of the electrons in the SSH and Holstein models. Note that the
excitons are hardcore as there cannot be more than one exciton per
site (there is either an exciton associated with a site, or there is
not).

Terms in the Hamiltonian (\ref{eqn:effhamiltonian}) have the following
origin: The first term $(E_2 + E_4)$ is simply an onsite, constant
offset of the energy which would be absorbed into a chemical
potential. It arises because of the interaction between the lasers and
the atoms, and would be present even in the absence of dipole-dipole
interactions between the atoms. The energy offset term $W$ arises
because the local dipole moment induces moments in other atoms, which
then cause an energy shift by re-interacting with the original
dipole. The most important term here is the long range hopping term
$\tilde{V}_{ij}$ which is the origin of a Su--Schrieffer--Heeger
interaction as discussed below. The corrections to this interaction
also have their origin in induced dipoles.

The following discussion is for arbitrary dimension and works equally
well for optical lattices in planes and on chains, and for phonons
that can vibrate on a line, plane or within a full 3D space. Phonons
can be introduced to the system by displacing the atoms from their
equilibrium positions. Atomic displacements within the optical lattice
are independent as no interaction is carried through the optical
lattice. Therefore, the vibrations of the atoms in the lattice are a
faithful representation of Einstein phonons. One interesting
possibility for 2D or 3D vibrations is to distort the harmonic
potential of the optical lattice to lift the degeneracy of the phonon
modes. The Hamiltonian for these modes will be,
\begin{equation}
H_{\rm ph}=\sum_{\nu\kvec}\hbar\omega_{0\nu}(d^{\dagger}_{\nu\kvec}d_{\nu\kvec}+1/2)=\sum_{\nu i}\hbar\omega_{0\nu}(d^{\dagger}_{\nu i}d_{\nu i}+1/2).
\end{equation}

The introduction of small in-plane phonon displacements, $\uvec_i$, at
site $i$ causes the interaction between Rydberg states to become,
\begin{equation}
V_{kl} = -\frac{\mu^2}{|\Rvec_k+\uvec_k -\Rvec_l-\uvec_l|^3},
\end{equation}
which can be Taylor expanded if the displacement is small, noting that
it is only necessary to expand to 1st order in $u$ as this is the
lowest order for non-trivial interaction with phonons. This leads to:
\begin{equation}
V(\Rvec+\uvec) \approx \frac{\mu^2}{|\Rvec|^3}-\frac{3\mu^2 \uvec\cdot\hat{\Rvec}}{|\Rvec|^4}.
\end{equation}

This expansion can be applied to the off-site terms in equation
\ref{eqn:dipoledressed} to obtain,
\begin{eqnarray}
\tilde{V}_{ij}(\uvec)& \approx &\frac{\alpha_S^2\alpha_P^2\mu^2}{1-\bar{\mu}^4/R_{ij}^6}\left[\frac{1}{R_{ij}^3}-\frac{3\uvec_{ij}\cdot\hat{\Rvec}_{ij}}{R_{ij}^{4}}\right.\nonumber\\
& & \left.-\frac{6\uvec_{ij}\cdot\hat{\Rvec}_{ij}}{R_{ij}^{10}}\frac{\bar{\mu}^4}{1-\bar{\mu}^2/R_{ij}^6}+\mathcal{O}(u^2)\right]
\end{eqnarray}
where $\bar{\mu}^2=\mu^2/(\Delta_S+\Delta_P)$ and $\uvec_{ij}=\uvec_{i}-\uvec_{j}$.

Quantisation of the atomic displacements using equation
\ref{eqn:quantise} leads to an extended Su--Schrieffer--Heeger (SSH)
model which is general for both cold ion or cold atom systems,
\begin{eqnarray}
H_{\rm ext-SSH} & = \sum_{ij}a^{\dagger}_i a_j&\frac{\alpha_S^2\alpha_P^2\mu^2}{1-\bar{\mu}^4/R_{ij}^6}\frac{1}{R_{ij}^3}\left[ 1- \frac{3}{R_{ij}}\sum_{\nu\kvec}\sqrt{\frac{\hbar}{2NM\omega_{\kvec\nu}}}\hat{\Rvec}_{ij}.\xivec_{\kvec\nu} \right.\nonumber\\
& &\left.
\left(d_{\kvec,\nu}[e^{-i\kvec\cdot\Rvec_{j}}-e^{-i\kvec\cdot\Rvec_{i}}]+d^{\dagger}_{\kvec,\nu}[e^{i\kvec\cdot\Rvec_{j}}-e^{i\kvec\cdot\Rvec_{i}}])\right)\right].
\label{eqn:sshgeneral}
\end{eqnarray}
Here, $a^{\dagger}_i$/$a_i$ create/annihilate excitons on site $i$ and
$M$ is the mass of the atoms (or ions).

A further simplification can be made for atoms in the optical lattice
because the oscillators are independent and local, thus there is no
momentum dependence in the phonon modes and
$\omega_{\kvec\nu}=\omega_{0\nu}$ and $\xivec_{\kvec\nu}=\xivec_{0\nu}$
so that the momentum sum only applies to the phonon creation and
annihilation operators. Moreover, the set of $\xivec_{0\nu}$ is now
orthogonal. Fourier transforming these operators, the local nature of
the model with Einstein modes becomes clear,
\begin{eqnarray}
H_{\rm ext-SSH} & = & \sum_{ij}a^{\dagger}_i a_j\frac{\alpha_S^2\alpha_P^2\mu^2}{1-\bar{\mu}^4/R_{ij}^6}\frac{1}{R_{ij}^3}\\&&\times\left[ 1 - \frac{3}{R_{ij}}\sum_{\nu}\sqrt{\frac{\hbar}{2M\omega_{0\nu}}}\xivec_{0\nu}\left(d_{j\nu}+d^{\dagger}_{j\nu}-d_{i\nu}+d^{\dagger}_{i\nu}\right)\right]\nonumber
\end{eqnarray}
This Hamiltonian is an extended version of the Su-Schrieffer-Heeger
model. Note that there can still be several phonon modes since even if
the atoms are arranged in a chain or sheet, the harmonic traps can be
three dimensional. Making the comparison with the SSH Hamiltonian
(Eq. \ref{eqn:ssh}), it is possible to assign values to $t_0$ and
$g$. The form of the coefficients mean that the relative strength of
the parameters can be tuned by changing the inter-ion spacing in the
lattice since the Hopping term, $t$, has $1/R^3$ functionality, whereas the
electron-phonon coupling term $g\sim 1/R^4$. The phonon frequencies
can be changed by altering the depth of the optical lattice, giving
full control over all parameters in the model. We will discuss this tuning in more detail in section \ref{sec:preparation}.

Making a similar Taylor expansion, phonon interactions via the on-site term in equation \ref{eqn:dipoledressedshift}
can also be considered,
\begin{eqnarray}
\sum_{k\neq j}\frac{1}{1-\bar{V}_{kj}^{2}}& \approx &\sum_{k\neq j}\frac{1}{1-\bar{\mu}^4/R_{kj}^6+6\uvec_{kj}\cdot\hat{\Rvec}_{kj}\bar{\mu}^4/R_{kj}^7}\\
& = & \sum_{k\neq j}\frac{1}{1-\bar{\mu}^4/R_{kj}^6}\left[1-\frac{6\uvec_{kj}\cdot\hat{\Rvec}_{kj}\bar{\mu}^4}{R_{kj}^7[1-\bar{\mu}^4/R_{kj}^6]}\right]
\end{eqnarray}
so that the local Hamiltonian becomes,
\begin{eqnarray}
H_{{\rm ext-Hol}}(\uvec) & = &\sum_{i}a^{\dagger}_{i}a_{i}(\Delta_S+\Delta_P)\alpha_S^2\alpha_P^2 \nonumber\\
& & \times \sum_{k\neq i}\frac{1}{1-\bar{\mu}^4/R_{ki}^6}\left[1-\frac{6\uvec_{ki}\cdot\hat{\Rvec}_{ki}\bar{\mu}^4}{R_{ki}^7[1-\bar{\mu}^4/R_{ki}^6]}\right]
\end{eqnarray}
the first term in this expression leads to a constant energy term. Once quantized, the second term leads to an extended Holstein type electron-phonon interaction. However, the prefactor $1/R^7$ indicates that this term will be very small in comparison to the SSH style term.

\begin{figure}
\begin{indented}
\item[]\includegraphics[width=85mm]{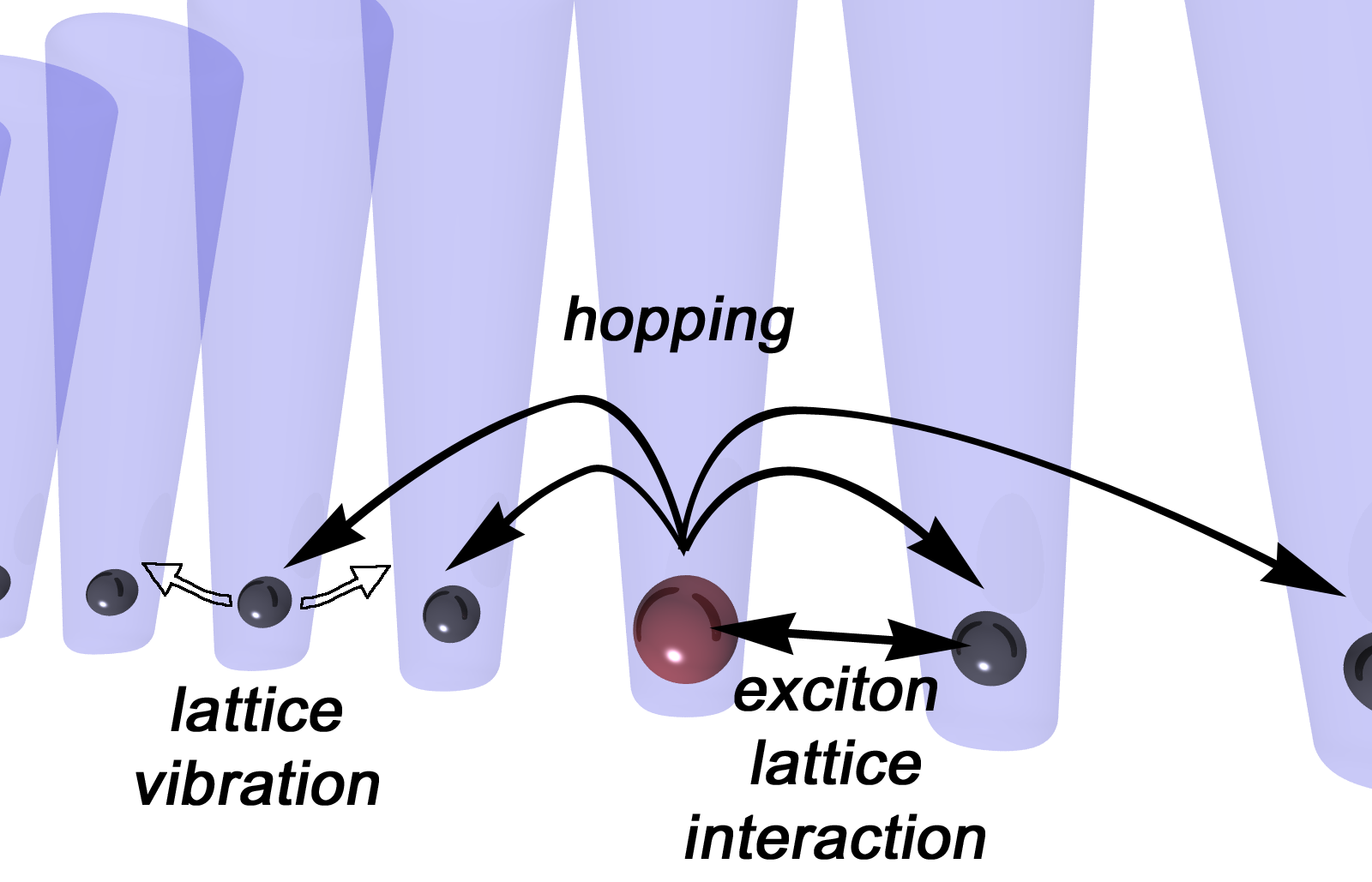}
\end{indented}
\caption{System of cold Rydberg ions / atoms used to simulate a polaron, annotated with Hamiltonian terms.}
\label{fig:simulator}
\end{figure}

Combining all Hamiltonian terms from the preceding discussion, the following Hamiltonian is found (up to a constant term):
\begin{equation}
H = H_{\rm ext-SSH}+H_{\rm ext-Hol}+H_{\rm ph}
\end{equation}
A schematic of the simulator with the effective interactions is shown in figure \ref{fig:simulator}.

\begin{figure}
\begin{indented}
\item[]\includegraphics[width=100mm]{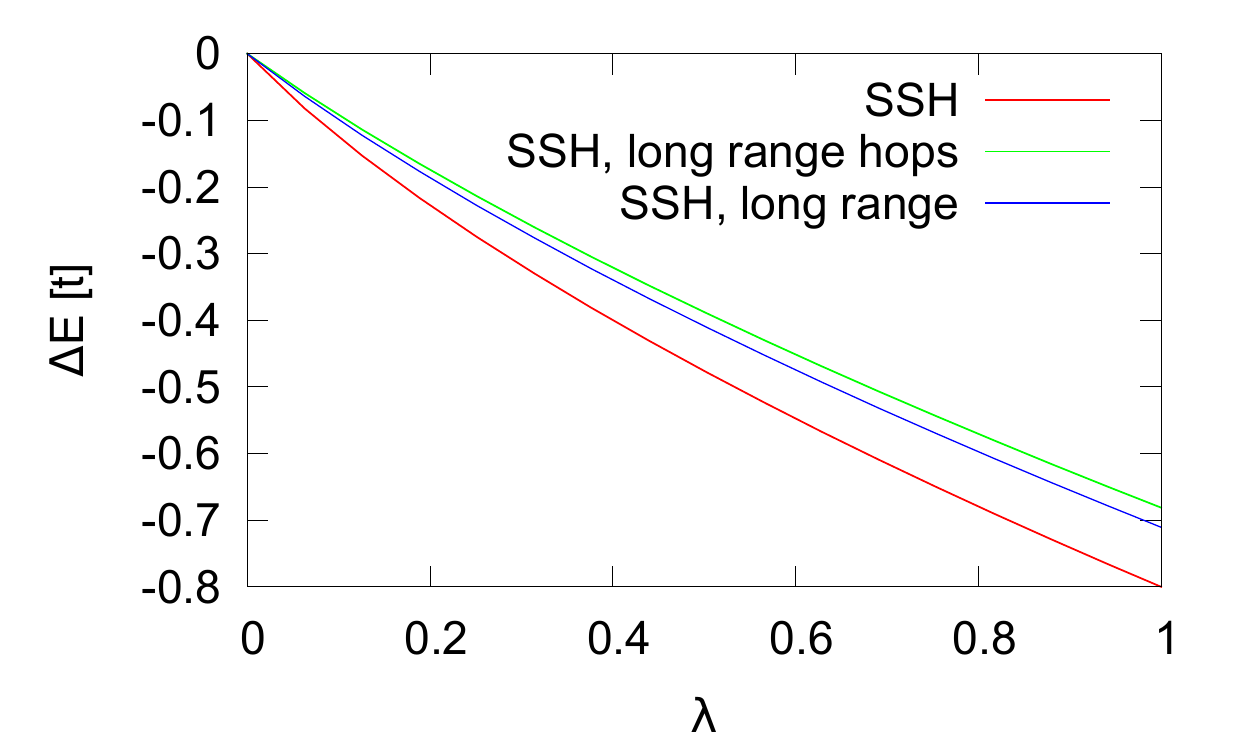}
\end{indented}
\caption{Effect of long range hopping and vertex on the polaron shift
  of the SSH polaron formed from interaction with one Einstein
  mode. Here $\lambda=g^2/\omega_0 t_0$ is the dimensionless electron-phonon coupling and $\omega_0=0.1t$. There is no qualitative change in the
  polaron shift as long range effects are switched on.}
\label{fig:polaronshift}
\end{figure}

\begin{figure}
\begin{indented}
\item[]\includegraphics[width=100mm]{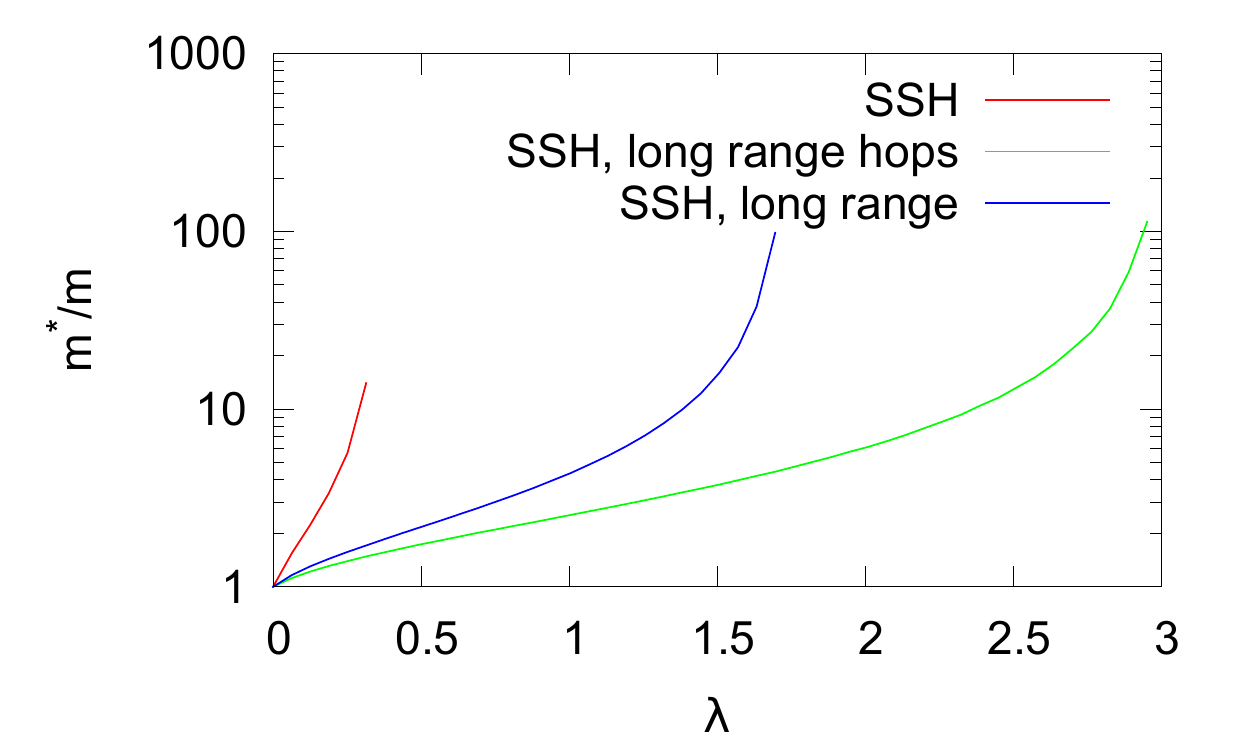}
\end{indented}
\caption{Effect of long range hopping and vertex on the effective mass of the SSH polaron formed from one Einstein mode.}
\label{fig:polaronmass}
\end{figure}

In order to gain insight into any differences that long range hopping
make to the extended SSH Hamiltonian, we make perturbation theory
calculations for three cases: The standard SSH model, an extended SSH
model with $1/R^3$ hopping and the extended SSH model with both
$1/R^3$ hopping and $1/R^4$ interaction strength. In all cases, the
model is on the chain. A single Einstein phonon mode in one-dimension
is considered. The lowest order perturbative contribution to the
self-energy of the polaron at absolute zero is:
\begin{equation}
\Sigma(\omega,\kvec) = \frac{\lambda}{2\pi}\int_{-\pi}^{\pi} dq \frac{g^2_{k,k+q}}{\omega-\omega_0-\epsilon_{k+q}+i\delta}
\end{equation}
where an integration over the energy has been performed for the phonon
propagator.  The Fourier transform of the interaction
vertex is,
\begin{equation}
g_{k,k+q}=2ig\sum_{n=1}^{\infty}\frac{1}{(R_n/R_1)^4}\left [\sin(n(k+q))-\sin(nk)\right]
\end{equation}
and the non-interacting dispersion,
\begin{equation}
\epsilon_{k} = -2t\sum_{n=1}^{\infty}\frac{\cos(nk)}{(R_n/R_1)^3}.
\end{equation}
The dimensionless electron-phonon coupling
$\lambda=g^2/\omega_0 t_0$ and $\delta$ is a small real number. From
this, the polaron energy can be calculated through the expression,
\begin{equation}
\omega-\epsilon_{k}-{\rm Re}\Sigma(\omega,k) = 0
\end{equation}
and the effective mass from the second derivative of the polaron
energy with respect to momentum, $k$.

Results from the perturbation theory are shown in
Figs. \ref{fig:polaronshift} and \ref{fig:polaronmass}. The polaron
shift shown in Fig. \ref{fig:polaronshift} is the difference in energy
between the interacting and non-interacting cases. For all three
cases, the form of the polaron shift is qualitatively
similar. Introduction of long range hopping decreases the polaron
shift as the polaron is essentially less tightly bound to a site and
hopping becomes easier, whereas introducing the long-range interaction
moderately increases the shift because the effective interaction is
slightly stronger. A stronger change can be seen in the effective mass
as the hopping range is increased (Fig. \ref{fig:polaronmass}) but the core physics remains the same. The
divergence in the effective mass signifies the breakdown in the
perturbation theory.

We note that the Rydberg excitation is Bosonic. For polaron problems,
which have a single electron, this is no problem because the electron
has no other particle with which to exchange. It is also possible
to simulate the ground state of a bipolaron (two interacting polarons)
since the ground state of two interacting Fermions is always a singlet
with symmetric spatial component of the wavefunction, with the
antisymmetry of the wavefunction imposed by the spin degrees of
freedom \cite{landauqmech}. Thus, most observables in the absence of a
magnetic field (e.g. total energy, effective mass) are identical in
the two Fermion or two Boson ground states. Since there can only be one exciton per site, the Bosons are hard core.

\section{Cold Rydberg Ions: Interactions with acoustic phonons}

An alternate possibility for tuning the Hamiltonian is to use cold
ions instead of cold atoms. For a system of cold ions confined to 1D,
the ions form a chain, held in place by the direct Coulomb repulsion
between the ions. Cold ions confined to 2D will form a Wigner crystal
with a triangular lattice. The phonon modes have been calculated in
the harmonic approximation for the triangular lattice
\cite{meissner1976a,bonsall1977a} and have an acoustic nature. Note
that these modes are not independent and the full Hamiltonian
\ref{eqn:sshgeneral} applies here. The ions considered here are
sufficiently heavy to justify the harmonic approximation. Therefore,
by changing from rubidium atoms to strontium ions, systems with
acoustic phonons are accessible.

Interaction between the Rydberg states and phonons occurs where there
are interactions terms between the dipole moments. For a system of
cold Rydberg ions with small dipole moment, the interaction potential
is \cite{mueller2008a},
\begin{eqnarray}
V(\Rvec_{i},\Rvec_{j},\rvec_{i},\rvec_{j}) & = & \frac{1}{R_{ij}}+\frac{(\Rvec_{i}-\Rvec_{j})\cdot(\rvec_{i}-\rvec_{j})}{R_{ij}^3}\nonumber\\& &+\frac{r_{i}^2-3(\nvec_{ij}\cdot\rvec_{i})^2+r^{2}_{j}-3(\nvec_{ij}\cdot\rvec_{j})^2}{2R_{ij}^{3}}\nonumber\\& &+\frac{\rvec_{i}\cdot\rvec_{j}-3(\nvec_{ij}\cdot\rvec_{i})(\nvec_{ij}\cdot\rvec_{j})}{R_{ij}^3}
\end{eqnarray}
The system should be prepared in the state discussed at the beginning
of section \ref{sec:coldatoms}, which simplifies the interaction to,
\begin{equation}
V(\Rvec_{i},\Rvec_{j})=\frac{1}{R_{ij}}+\frac{\tilde{\mu}^2}{R_{ij}^3}
\end{equation}
The monopole ion-ion interaction is only responsible for the phonon
modes and the effective dipole-dipole interaction is identical to the
cold atoms case. Thus, beyond a change in the form of the phonons, the
resultant exciton-phonon interaction is the same for atoms and ions. We note that a fully excited $|nS\rangle$ ion moves in a different potential to an ion in a fully excited $|nP\rangle>$ state. Since the ions are dressed, they have only a very small admixture of the Rydberg states, so the motion of the ions is mainly determined by the ground state properties of the ions which are very similar for $|g\rangle$ and $|h\rangle$ states, and therefore it is safe to assume that all ions move in approximately the same potential.

\section{Experimental considerations}
\label{sec:preparation}

\begin{figure}
\begin{indented}
\item[]
\includegraphics[width=10cm]{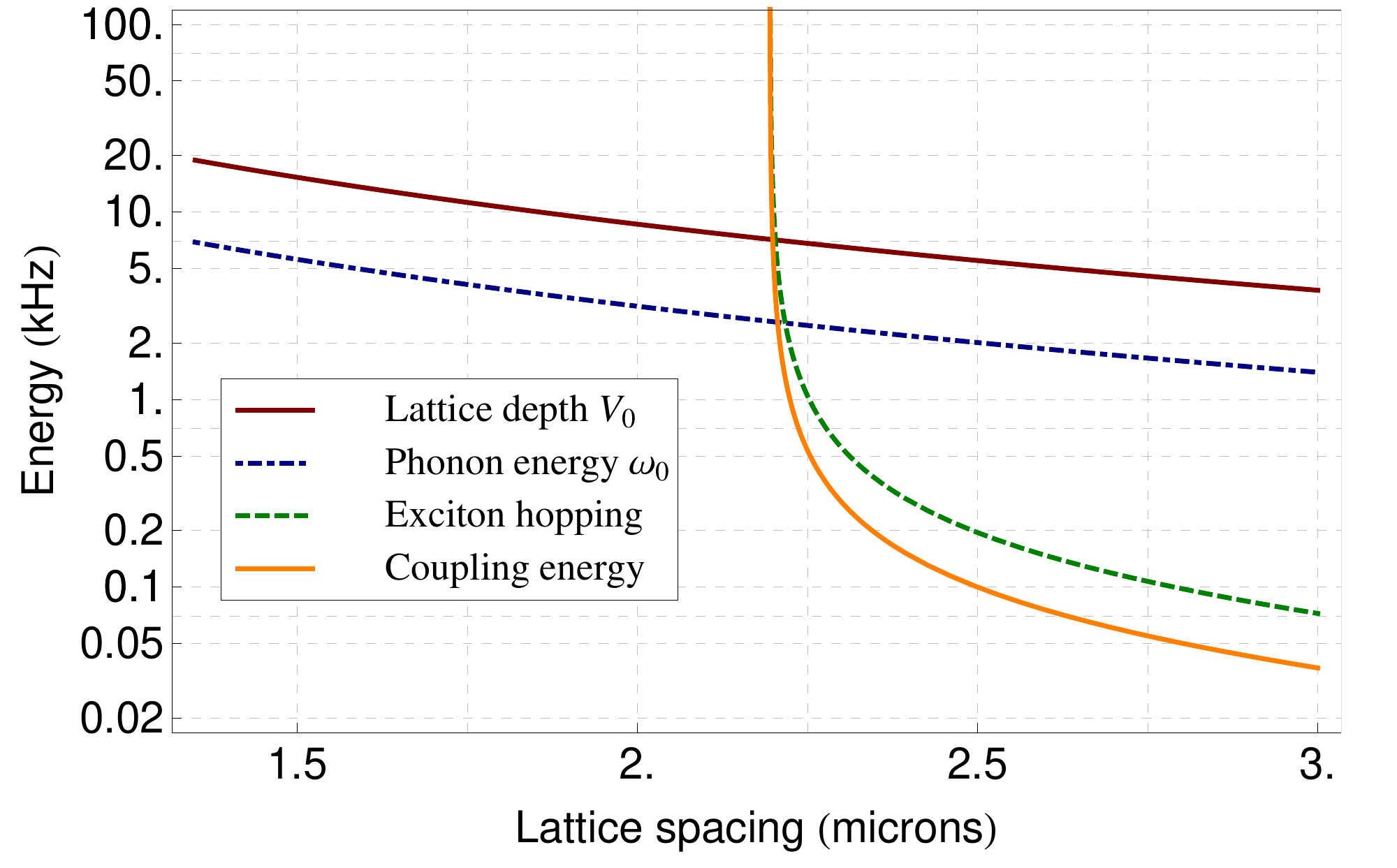}
\end{indented}
\caption{Relative strength of on-site and near-neighbour energies in an optical lattice. $\Omega=10$MHz and $\Delta=100$kHz. The polarons formed from weak coupling with high-energy phonons can be investigated.}
\label{fig:tunabilitylattice}
\end{figure}

So far we have discussed the Hamiltonian of the atomic/ionic system and its condensed matter analogues. Now we sketch the manner in which experiments may be performed.

In the case of atoms or ions, we assume that each lattice site is occupied by a single atom/ion, and each site may be addressed individually in such a way as to allow the preparation of the initial simulator state $|g\rangle$ or $|h\rangle$ on each atom/ion. We assume also that the state of each atom may be measured by use of a laser tuned close to resonance with a closed transition involving the ground state of interest. Thus, the state of each atom/ion is detected in fluorescence.

In the case of atoms, the system may be prepared from a Mott insulator phase where a single atom occupies each lattice point. Alternatively, there exist proposals to obtain single occupancy in non-degenerate (and hence much lower density) atomic ensembles \cite{beterov2011}. For ions at low temperatures a suitable lattice is necessarily formed; it is possible to cool the motional states of the ion crystal using finely tuned lasers \cite{king1998}.

First, we discuss the Rydberg states which cause the atom-atom
interaction in our SSH model. We couple our two hyperfine ground
states $|g\rangle$, $|h\rangle$ to the $|nS\rangle$ and $|nP\rangle$
Rydberg state, respectively. In the following we discuss the atom
Rb$^{87}$. These particular Rydberg states interact according to the
dipole-dipole Hamiltonian in Eq. \ref{eqn:dipdipint}. In detail, this
takes the form:
\begin{equation}
V = \frac{1}{4\pi \epsilon_{0}}\sqrt{\frac{8\pi}{3}}\frac{(d_{nS;nP})^2}{R^3}(-1)^{mJ}\langle 1,m;1,m^{\prime}|2,m-m^{\prime}\rangle Y_{2, m-m^{\prime}}(\Rvec)
\end{equation}
where the vector connecting the two atoms is $\Rvec$, $d_{nS;nP}$ the
dipole matrix element between the states $|nS\rangle$ and $|nP\rangle$
and $Y_{2, m-m^{\prime}}(\Rvec)$ is a spherical harmonic. The lasers
select only the $m=0$ and $m^{\prime}=0$ magnetic sublevels. In
general, the dipole-dipole interaction scales with $n^{4}/R^{3}$ and
the lifetime of the Rydberg states scales with $n^{3}$, which together
mean that choosing the highest energy Rydberg state possible leads to
the best possible properties of the atoms and ions for the
simulator. However, experimentally we stress that for a given laser
intensity, the Rabi frequency of the dressing laser scales with
$n^{-3/2}$, and the shift of the $|nS\rangle$ atoms due to their
mutual interaction (the ``blockade shift'') scales with
$n^{11}/R^6$. Thus one cannot go to arbitrarily high $n$ without
shifting the $|nS\rangle$ states too far from the dressing laser, or by making
unreasonable demands on the laser intensity. We choose the $n=48$
states for which the laser power required to drive the $|g\rangle
\rightarrow |nS\rangle$ and $|h\rangle \rightarrow |nP\rangle$
transition should be reasonable (the Rabi frequency for these
transitions scales with $n^{-3/2}$), but where the 48S - 48P
interaction is 1252 MHz$/\mu$m$^{-3}$; large enough for our
purposes. The lifetimes of these states are: $\tau_{48S}=106 \mu$s and
$\tau_{48P}=117 \mu$s, but the effective lifetime of the dressed
states are lengthened by a factor $\alpha_{S}^{-2}$ and
$\alpha_{P}^{-2}$. Using this expression for $V$ and reasonable values
for the Rabi frequencies $\Omega_{S}=\Omega_{P}=$10 MHz and detunings
$\Delta_{S}=\Delta_{P}$=100 MHz we calculate the nearest neighbour
exciton hopping rate $\tilde{V}_{ij}$ for a separation of $2.5$ micron
to be 0.19 KHz, and a dressed state lifetime of 48 ms $\approx 54$
exciton hopping times. We note that for these choices of
$\Delta_{S}$,$\Delta_{P}$ and $\Omega_{S}$, $\Omega_{P}$,
$\tilde{V}_{ij}$ diverges below a separation of about 2.2$\mu$m, which
sets an additional constraint on the types of experiments we can
perform. We also know that the $|nS\rangle$ atoms will interact with each other
due to the van der Waals interaction $V_{vdW}=C_{6}/R_{ij}^{6}$, where
$C_{6}$ is the appropriate van der Waals coefficient. For our $48S$
states, $C_{6}\approx 1050$ MHz$/\mu$m$^{6}$.

Now we turn to considering some of the experimental values required for
modelling an SSH Hamiltonian. We have in mind two set ups which illustrate the utility of the model and some of the experimental issues that must be addressed. In the first case, a linear chain of
atoms (of mass M) is confined by a far-detuned optical lattice formed from
a pair of intersecting laser beams with wavelength $\lambda$. As mentioned above we stipulate that there be a single atom per lattice site. With counter-propagating lasers, such a
system is characterised by the recoil energy $E_{r}=2 \pi \hbar^{2}/
M \lambda$ and the potential depth $V_{0}$, which depends on the laser intensity. Assuming that the atoms are trapped near the potential minima, the phonon frequency will be $\omega_{0}=2E_{r}\sqrt{ V_{0}/E_{r}}/\hbar$, the atomic hopping energy $J=(4/ \sqrt{\pi})E_{r}(V_{0}/E_{r})^{3/4}e^{\sqrt{V_{0}/E_{r}}}$ and we estimate the on site atomic repulsion energy as $U=\sqrt{8/\pi}( a_{s}/a_{0})E_{r}(V_{0}/E_{r})^{3/4}$, with $a_{s}$ the s-wave scattering length for the atoms - 5.3 nm for Rb$^{87}$, and $a_{0}$ the length of the ground state wavefunction.
In order to initialise the system, it is necessary that the atoms be individually addressable,
which determines the minimum lattice wavelength to be around 1
$\mu$m, however we adopt the more conservative 2.25 $\mu$m lattice wavelength considered above. In order that atomic tunneling rate $J$ be suppressed, we require that $V_{0} >>20 E_{r}=4.53$ kHz;
for the 2.25 $\mu$m lattice spacing, $\tilde{V}_{ij}=1.43$ kHz and
$\omega_0 = 2.03$ kHz. We then have $J \approx 0.001$ kHz, $U=2.3$ kHz, increasing the localisation further. We note that this phonon
frequency is a large fraction of the site depth $V_{0}$. Also, for atomic spacings of 2.5 microns and upwards, $V_{vdW}<24$ MHz, smaller than, but comparable to, the detunings $\Delta_{S}$ and $\Delta_{P}$ we have chosen for our example.

The tunability of the cold atom system in an optical lattice where the
atomic potential $V_{latt}=V_0 \sin^{2}(kx)$, $V_{0}=20E_{r}$ is illustrated in
Fig. \ref{fig:tunabilitylattice} as a function of the lattice wavelength. The figure
shows the relative energies associated with the trap depth, the
phonons, the nearest-neighbour exciton-phonon coupling and the exciton
hopping. Using realistic parameters, the phonon energy and coupling
energy can be swapped. The resulting SSH model is in the weak
coupling, high phonon frequency regime.

\begin{figure}
\begin{indented}
\item[]
\includegraphics[width=10cm]{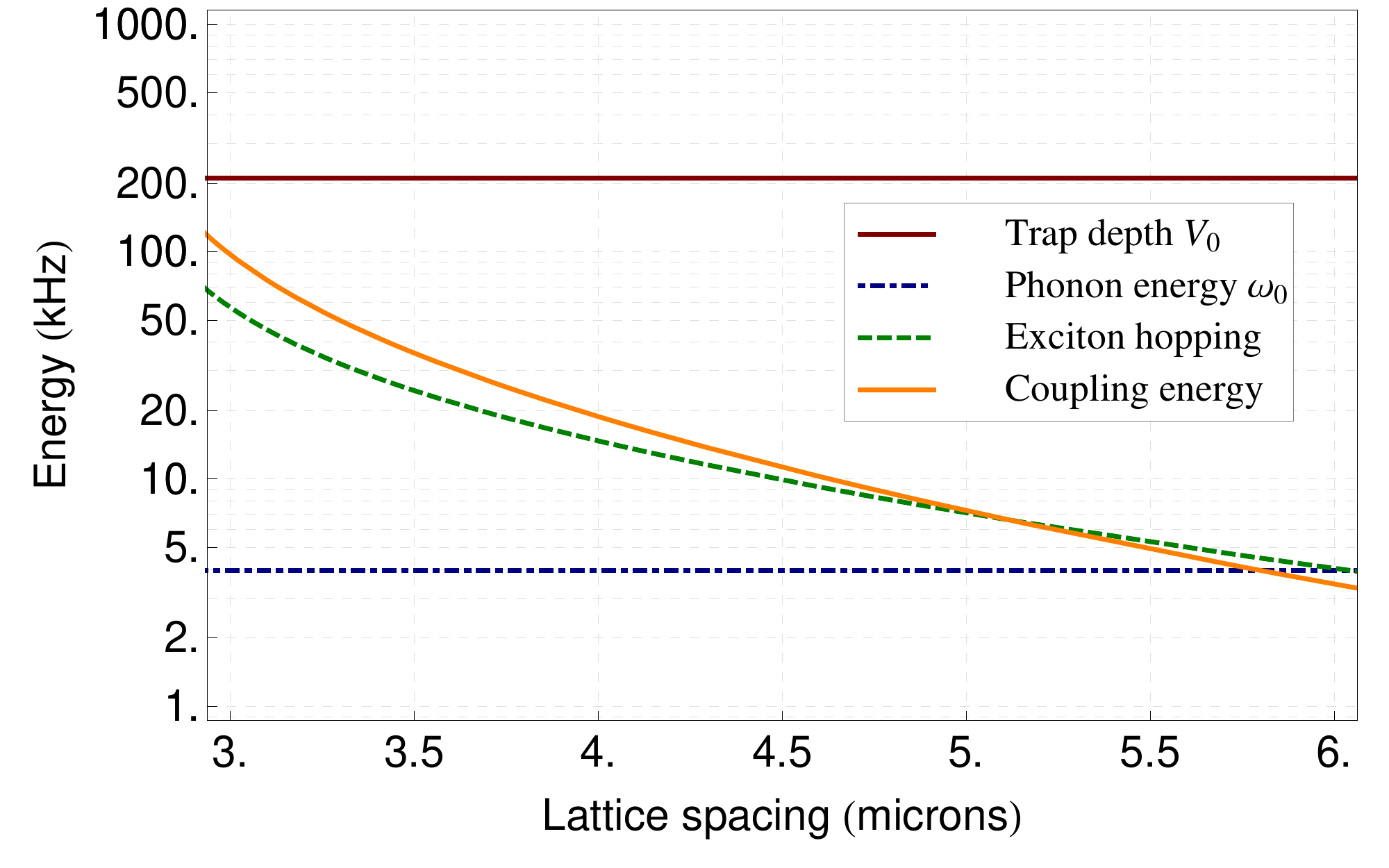}
\includegraphics[width=10cm]{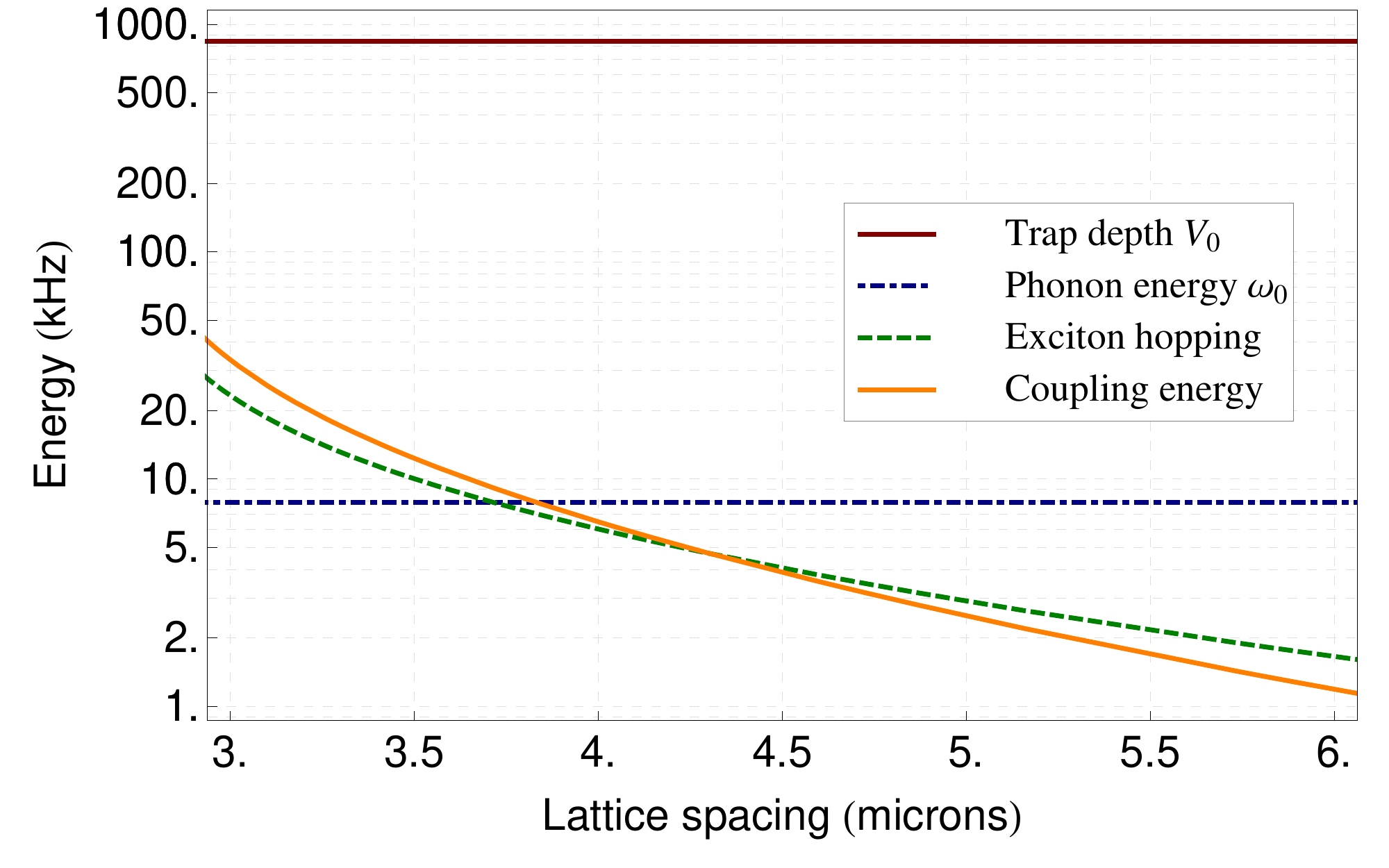}
\end{indented}
\caption{Relative strength of on-site and nearest-neighbour energies in painted potentials. All hierarchies of energies can be realised. The parameters here are detuning of $\Delta=100kHz$, Rabi frequency $\Omega=25$MHz and the waist is $2.5\mu$m. (top) The painted potential is made with a $P=0.75$mW laser, and the exciton hopping and coupling energies cross above the phonon energy so that the regimes where phonon energy $<$ hopping $<$ coupling and phonon energy $<$ coupling $<$ hopping can be explored. (bottom) $P=0.75$mW, and the exciton hopping and coupling energies cross above the phonon energy so that the regimes of phonon energy $>$ hopping $>$ coupling and phonon energy $>$ coupling $>$ hopping can be explored.}
\label{fig:tunabilityspot}
\end{figure}

In the next experimental scenario we consider how a far more tunable
system can be constructed by using painted potentials
\cite{Henderson2009}. In this system the lattice potential is an array
of Gaussian laser spots, with power $P$ and focused to a waist
$w_{0}$, in which case the traps have a potential $V_{trap}=V_{Opt}
e^{-2r^2/w_{0}^{2}}$ where the trap depth is $V_{Opt}=2P/\pi
w_{0}^{2}$ and the phonon frequency is $\omega_{0}=(1/w_{0})\sqrt{V_{Opt}
  /M}$. The trap spacing, $d$, may be chosen such that tunneling is
negligible by choosing $d>w_{0}$, and the phonon frequencies may then
be determined by the laser power $P$. In this case, we need to choose
higher Rabi frequencies and detunings $\Omega_{S,P}=20$ MHz and
$\Delta_{S,P}=100$ MHz respectively. For these parameters, and with
the larger atomic separations ($>4 \mu$m), we calculate that the
dressed state lifetimes are $\approx 12$ ms, which is about 85
$\tau_{hop}$, and that the blockade shift of the $nS$ states is less
than 2MHz. We plot the the parameters associated with this potential
in Fig. \ref{fig:tunabilityspot}. Here, all hierarchies of phonon
energy, coupling constant and exciton hopping can be achieved. A major
advantage of the painted potentials is that much deeper potentials can
be achieved, but phonon frequencies can be much smaller, putting the
effective model in the physical regime of a condensed matter
system. Also, rings of atoms can be created to explore periodic
boundary conditions.

For modelling the SSH model using cold ions, we have not undertaken
any detailed calculations to assess our simulator proposal, but we can
make some points here. Ions trapped in an ion trap have phonon
frequencies in the 100 kHz - 2 MHz range, and spacing on the order of
several $\mu$m. Thus, the phonons are very much higher in energy than
the atomic case. Recent studies of Rydberg states of ions has shown
that the interactions between the ions are very weak - additional
complications involving RF dressing of the rydberg atoms are required
\cite{Kaler2011a}. This means that implementation of the cold ion
extension to simulate polarons formed from acoustic phonons in regimes
relevant to condensed matter physics will be a major experimental
challenge.

We end by describing an experiment to explore the self trapping regime
of the SSH model which is a subject of current interest
\cite{marchand2010a}. We consider studying the phenomena of
self-trapping in the SSH model on a circular array of Gaussian spots -
that is, a ring of traps. A particular lattice point is chosen for the
location of a single excitation, and then the dressing fields are
rapidly switched on. Since this impulse contains multiple momenta, the
localized exciton state will expand into the lattice. The result of
the simulation may be obtained by extracting the location of the
excitation as a function of time. This may be achieved by rapidly
switching off the dressing potentials, and then measuring the state of
the atoms/ions at each lattice point. In the self-trapped regime, the
exciton hopping will be suppressed which will be observed as a
narrowed distribution of final exciton locations. The hopping leads to
a quantum random walk \cite{kempe2003a} so the mass can be extracted from the
spatial distribution of an ensemble of measurements. Note that there
are significant differences between the distributions arising from
classical and quantum random walks.

\section{Summary and outlook}
\label{sec:summary}

In this article, we have shown how systems of cold Rydberg atoms and
ions can be used as a simulator for electron-phonon interactions, and
derived the mapping onto a Su-Schrieffer-Heeger model, which is used
to simulate strongly deformable materials. We have also discussed how
readout of physically relevant properties could be carried
out. Further, we have discussed the parameters of the model that can
be achieved, showing that rubidium atoms in painted potentials can be
used to access physically relevant couplings, phonon frequencies and
exciton hoppings.

It is relevant to discuss scalability here. A drawback of the current
approach is that the excitations are Bosons rather than Fermions. It
may be possible to simulate full fermion-phonon interactions using a
more complicated two layer 2D optical lattice with dressed Rydberg
atoms in the lower layer and an incomplete layer of Rydberg atoms in
the upper layer. The Rydberg atoms in the upper layer will interact
with those in the lower layer through the dipole-dipole interaction,
leading to a Hubbard-Fr\"ohlich model, but with the caveat that the
system is much more complicated \cite{hague2011b}. We note that much
has been learnt by simulating Bose-Hubbard models with cold atoms, so
we consider that the lack of Fermionic excitations is currently a
minor drawback. We hope that the proposed system will stimulate
experimental work in the area, and encourage the development of
quantum simulators for electron-phonon interactions which work with
Fermionic excitations. In the case of polymer physics, the quantum
simulations are relevant as they may shed light on ongoing debates
about self-trapping phenomena \cite{marchand2010a}.

\ack

JPH would like to thank EPSRC grant no. EP/H015655/1 and CM grant
no. EP/F031130/1 for financial support. We would like to thank
I. Lesanovsky, M. Bruderer, A. Kowalczyk, N. Braithwaite,
S. Bergamini, P. Kornilovitch, S. Alexandrov and J. Samson for useful
discussions.

\section*{References}
\bibliography{coldionphonons_ssh}

\end{document}